\documentclass[superscriptaddress,twocolumn,showpacs,prb]{revtex4-1}
\usepackage[utf8]{inputenc}
\usepackage{braket}
\usepackage{graphicx}
\usepackage{amsfonts}
\usepackage{pgfplots}
\usepackage{csquotes}
\usepackage{hhline}
\usepackage{amssymb}
\usepackage{listings}
\usepackage{color}

\definecolor{codegreen}{rgb}{0,0.6,0}
\definecolor{codegray}{rgb}{0.5,0.5,0.5}
\definecolor{codepurple}{rgb}{0.58,0,0.82}
\definecolor{backcolour}{rgb}{0.95,0.95,0.92}
\lstdefinestyle{mystyle}{
    backgroundcolor=\color{backcolour},   
    commentstyle=\color{codegreen},
    keywordstyle=\color{magenta},
    numberstyle=\tiny\color{codegray},
    stringstyle=\color{codepurple},
    basicstyle=\footnotesize,
    breakatwhitespace=false,         
    breaklines=true,                 
    captionpos=b,                    
    keepspaces=true,                 
    numbers=left,                    
    numbersep=5pt,                  
    showspaces=false,                
    showstringspaces=false,
    showtabs=false,                  
    tabsize=2
}

\usepackage{subfigure}

\usepackage{dcolumn}
\usepackage{tabularx}
\usepackage{caption}
\usepackage[a4paper,margin=2cm]{geometry}
\usepackage{graphicx}
\usepackage[colorlinks=true,linkcolor=blue,citecolor=blue,urlcolor=blue]{hyperref}
\usepackage{braket}
\usepackage{float}
\newcolumntype{C}{>{\centering\arraybackslash}X}

\begin{document}

\title{Quantum Go: Designing a Proof-of-Concept on Quantum Computer}
\author{Shibashankar Sahu}
\email{shibass19@iiserbpr.ac.in}
\affiliation{Indian Institute of Science Education and Research, Berhampur-760005}
\author{Biswaranjan Panda}
\email{biswaranjanp19@iiserbpr.ac.in}
\affiliation{Indian Institute of Science Education and Research, Berhampur-760005}
\author{Arnab Chowhan}
\email{arnab.chowhan@cbs.ac.in}
\affiliation{Centre for Excellence in Basic Sciences, University of Mumbai, Mumbai-400098}
\author{Bikash K. Behera}
\email{bikas.riki@gmail.com}
\affiliation{Bikash's Quantum (OPC) Pvt. Ltd., Balindi, Mohanpur 741246, West Bengal, India}
\affiliation{Department of Physical Sciences,\\ Indian Institute of Science Education and Research Kolkata, Mohanpur 741246, West Bengal, India}
\author{Prasanta K. Panigrahi}
\email{pprasanta@iiserkol.ac.in}
\affiliation{Department of Physical Sciences,\\ Indian Institute of Science Education and Research Kolkata, Mohanpur 741246, West Bengal, India}

\begin{abstract}
The strategic Go game, known for the tedious mathematical complexities, has been used as a theme in many fiction, movies, and books. Here, we introduce the Go game and provide a new version of quantum Go in which the boxes are initially in a superposition of quantum states $\Ket{0}$ and $\Ket{1}$ and the players have two kinds of moves (classical and quantum) to mark each box. The mark on each box depends on the state to which the qubit collapses after the measurement. All other rules remain the same, except for here, we capture only one stone and not chains. Due to the enormous power and exponential speed-up of quantum computers as compared to classical computers, we may think of quantum computing as the future. So, here we provide a tangible introduction to superposition, collapse, and entanglement via our version of quantum Go. Finally, we compare the classical complexity with the quantum complexity involved in playing the Go game.
\end{abstract}

\maketitle
\section{Introduction}
The door of quantum computing was opened by the works of Benioff \cite{qgo_Benioff1979}, Manin \cite{qgo_Manin1980}, and Feynman \cite{qgo_Feynman1982}, and the concept was raised by Jozsa \cite{qgo_DeutschJozsa1992}and others. Processing power and memory optimization are the key advantages of quantum computers, the impetus being entanglement and superposition \cite{Ent-Sup}. A limitation is quantum decoherence \cite{qgo_Franklin2004,qgo_MishraRG2020}, caused by the interference of qubits with the surroundings. Still quantum computers are thought to solve several complex problems and simulate quantum physics far better than classical counterparts.\\
Quantum game theory was developed \cite{qgo_JackHidary} from the works of Wiesner on quantum money \cite{qgo_Weisner1983}, followed by the works of Deutsch and Jozsa  on quantum information and game theory formalism by Meyer \cite{qgo_Meyer1999} and Eisert \cite{qgo_Eisert2000}. Lately, it has attracted much attention, be it quantum Tic-Tac-Toe (by Sagole \textit{et al.}) \cite{qgo_Sagole2019}, quantum Sudoku (by Pal \textit{et al.}) \cite{qgo_Pal2019}, quantum Go (by Ranchin) \cite{qgo_Ranchin2016}, quantum Chess (by Kartavicius) \cite{qgo_RyanKartavicius}, quantum Pong (by Verma \textit{et al.}) \cite{qgo_VermaRG2019}, quantum Bingo (by Singh \textit{et al.}) \cite{qgo_SinghRG2019}, quantum Monty Hall (by Paul \textit{et al.}) \cite{qgo_PaulRG2019} or quantum Diner's Dilemma game (by Anand \textit{et al.}) \cite{qgo_AnandRG2019} to name a few.\\
In today's era of artificial intelligence, deep learning methods and artificial neural networks have made tremendous developments. We can remark the example of the dominance of computer programs over human players in Go game: DeepMind technologies developed AlphaGo, which became the first Go program to beat a professional Go player on 19 $\times$ 19 board. Its playing style is having more probability of victory by fewer points rather than less winning probability by more points. It was succeeded by three more powerful programs, namely AlphaGo Master, AlphaGo Zero, and AlphaZero. To make it more challenging for the computers to master, we have provided a novel variant of the quantum Go game and simultaneously tried to provide the reader an interesting exposure to some basic quantum phenomena such as superposition, collapse, and entanglement.\\
IBM quantum experience (IBM QE), a superconducting quantum computing system provides user-friendly access to the quantum computers to the researchers all around the globe. Many important applications have been found by IBM QE \cite{application1,application2,application3,application4,application5,application6,application7,application8}. In this paper, a 16-qubit quantum simulator and a 5-qubit quantum computer are used to create superposed states in the $4\times4$ quantum Go board. Then we show the circuits used during the first two moves in the game. We present the capturing of the stones used in the game through the quantum circuits. First, we proceed for $3\times3$ board for implicitly. We then design the quantum circuit diagram for capturing ($3\times3$ board) of the stones. Then a basic mathematical calculations of qubits which are needed for $4\times4$ board, is given. Finally, complexity of the game is calculated and is generalized for $n\times n$ board.

\section{Classical Go Game}
Like chess, Go game is a deterministic perfect information game where no information is hidden from either player, and there are no built-in elements of chance, such as dice. It is a two-player game, in which the aim is to surround more territories than the opponent. Go, probably the world's oldest board game is thought to have originated in China some 4000 years ago. According to some sources, this date is as early as 2356 BCE. The game was probably taken to Japan about 500 CE, and it became popular during the Heian period (794-1185). The modern game began to emerge in Japan with the subsequent rise of the Samurai class. It was given special status there during the Tokugawa period (1603-1867) when four highly competitive Go schools were set up and supported by the government and Go-playing was thus established as a profession. The game became highly popular in Japan in the first half of the 20th century; it was also played in China and Korea, and its following grew there in the latter decades of the century. Play spread worldwide after World War II.  More recently, it is being played electronically.

\subsection{Rules of Go}
\begin{figure}
\centering
\includegraphics[scale=0.7]{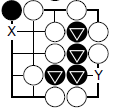}
\caption{The black stone is captured if white puts in `X' and all the black stones marked with triangles get captured if white puts in `Y'}
\label{qgo_Fig1}
\end{figure}

\begin{itemize}
\item \textbf{Basics:} The $13\times13$, $9\times9$, or $19\times19$ board starts empty. One player has black stones and the other has white. Black goes first, and then the players take turns in which a stone is to be put down on the board on the corners of the squares.

\item \textbf{The object of the game:} At the end of the game we score one point for each point of territory we have, and one point for each stone we have captured. The person with the most points wins.
    
\item \textbf{Territory:} If on moving from a white (or black) point along the lines of the board, we always come to the edge of the board or a white (or black) stone, never a black (or white) one. Then these empty points belong to white (or black). 

\item \textbf{Liberties:} The liberties of stone are the empty points that are next to it. A stone at the corner has 2 liberties, at the edge has 3 and others have 4. Liberties are important because if a stone runs out of them it will be captured!

\item \textbf{Chains:} If we put two of your stones next to each other, they become a small chain.

\item \textbf{Capturing:} The isolated black stone in Fig. \ref{qgo_Fig1} has only one liberty left, marked with an X. If we place a white stone at the position X then the black stone loses liberty and we can take the black stone off the board . Similarly we can capture a chain of stones. For example we can capture the chain of black stones marked with triangles just by placing a white stone at the liberty Y. After that we can take off the chain of four black stones.

\item \textbf{Suicide:} If we place a stone to form a chain without any liberty, it is a suicide in the game.
\item \textbf{When does the game end?} when our turn comes, we can just simply pass without making a move. If we pass, we give an extra stone to our opponent in order to add to their captures. A player does pass if there is further no chance to surround more territory or to attack  enemy stones. If both the players consecutively declare pass, then the game is over. In order to make same the number of the moves for both players, the white should pass last although it means the third pass in a row \cite{Ref11}.
\end{itemize}

Hence, the general strategy should be clear:

\begin{itemize}
    \item Try forming chains,
    \item Try gaining territories,
    \item Protect stones,
    \item Avoid suicides.
\end{itemize}

\section{Quantum Go Game}
This section gives a detailed description of our version of the quantum Go game. The game uses simple quantum computing principles, which makes it different and more interesting than the regular classical version of Go. The basic rules are almost the same with a major difference being the inclusion of probability in the classical game. In the game, a player chooses a particular box in his (or her) turn and the output is either Black or White with a 0.5 probability each. Another modification in our game is that only the capture of a single stone will be considered, and for simplicity, the capture of chains will be ignored. Let us look for our game version in a simple $4\times4$ board. Initially, the board is set up as Fig. \ref{qgo_Fig2}. Each box is in a superposition of states $\Ket{0}$ and $\Ket{1}$. It is a quantum state that has an equal probability of collapsing into a classical state, i.e., $\Ket{0}$ or $\Ket{1}$. The game allows two legal moves:

\begin{figure}[!h]
\centering
\includegraphics[scale=0.7]{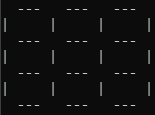}
\caption{4 $\times$4 quantum Go game board before the game starts.}
\label{qgo_Fig2}
\end{figure}

\begin{enumerate}
\item \textbf{Classical Move:} When applied to a box, it collapses to either of the classical states with equal probabilities.
\item \textbf{Quantum Move:} It requires a control box and a target box. It uses the concept of \textit{quantum entanglement} by entangling the target box with the control box in such a way that when a classical move is applied to the control box, both collapse to the same state.
\end{enumerate}

A \textit{classical move} cannot be applied to the same box twice. Quantum move's target box should be in a classical state, and the control box should be in a quantum state so that when the quantum state of the control box collapses to the favored state of the player, it reverses the classical state of the target box. Black player's favored state is $\Ket{1}$, and player white player's favored state is $\Ket{0}$. A box is marked `B' when it collapses to state $\Ket{1}$ and `W' when it collapses to state $\Ket{0}$. Let us look at an example of how the game works. Our two players are Bob (Black) and Alice (White). Bob takes the first turn which ought to be classical move as there is no box in the classical state for being the target for his quantum move. So, he chooses any one of the 16 boxes, and the chosen box collapses into either state $\Ket{0}$ or $\Ket{1}$. Suppose, he is lucky and the box collapses to $\Ket{1}$ state, which is shown by marking the box `B'. Now, Alice's turn: she may choose to go for a classical move and collapse any of the other boxes or may play a quantum move and entangle another box with the box marked `B', the latter being the target of the entanglement. Now, whenever a classical move is applied onto the control box and if it collapses to classical state $\Ket{0}$, it automatically reverses the state of the target, which in this case makes the previously `B' marked box `W'. But if the control collapses into $\Ket{1}$ state, then the target state would remain the same. The board and the number of stones captured after each step will be shown and after the game is over, the players can count their territory numbers by Rule-3 and the winner will be decided by Rule-8.

\section{Circuit Explanation}
The quantum circuit is set up with 16 qubits, and 16 classical bits and Hadamard gates are applied to each qubit. Hadamard gate creates the superposition state, which we call the quantum state. Let us say Bob chooses to go for a classical move on box 4, then as soon as the input is received the qubit associated with box 4 (i.e. qubit 3 as qubits count start from qubit 0) is measured and the result is analyzed. Box 4 is marked `B' if the result is $\Ket{1}$ and `W' if the result is $\Ket{0}$. To ensure, qubit 4 does not alter in subsequent moves, we add a Reset gate if it collapses to $\Ket{0}$; or a Reset followed by a Not gate if it collapses to $\Ket{1}$. Now it is Alice's turn and let us say she chooses to go for a quantum move. So, she entangles box 4 with let us say box 2 (i.e., qubit 3 with qubit 1), with box 4 being the target and 2 being the control. An anti-control NOT gate is applied, with qubit 1 being the control and qubit 3 the target. What this move does is, if qubit 1 is measured and it collapses to $\Ket{0}$, then the anti-control NOT gate would reverse the state of qubit 3 and the previously marked `B' would change to `W', which would ultimately make both box 2 and box 4 marked as `W'. On the contrary, if qubit 1 collapses to $\Ket{1}$, anti-control NOT gate would not be able to reverse the state of qubit 4, and both boxes would be marked `B'. The game continues similarly until both players agree to pass. An anti-control NOT gate is added if the white player goes for a quantum move and a control NOT gate is applied if the black player goes for a quantum move. The required circuits for the above explanation are shown on the next page.

\section{Building the game \cite{Ref12}}
\subsection{STEP-1}
\textit{Making the board: }
board()
\begin{itemize}
    \item Takes the input from the user for board size and print board of that size (as shown above for 4$\times$4).
\end{itemize}
\subsection{STEP-2}
\textit{Generating the circuit:}
setup()
\begin{itemize}
    \item Generates a circuit of \(n^2\) qubits and \(n^2\) classical bits, \(n^2\) being the board size.
    \item Adds Hadamard gate to all the qubits.
    \item Prints the rules of the game.
\end{itemize}
\subsection{STEP-3}
\textit{Notifying whose turn:}
turn()
\begin{itemize}
    \item Checks whose turn it is and prints accordingly.
    \item Asks the player for classical or quantum move and corresponding position and calls cmove() or qmove() accordingly.
    \item Else if the player chooses to pass, looks for the response of the other player and if all agree, ends the game, ensuring white passes last.
\end{itemize}
\subsection{STEP-4}
\textit{Classical move: }
cmove()
\begin{itemize}
    \item Calls the Aer() function from the qiskit library
    \item Simulates the circuit on qasm simulator
    \item Checks if the position is already marked
    \item Checks if the position is entangled with any other position due to previous moves
    \item If yes, adds CNOT or anti-CNOT as mentioned before
    \item Else collapses the qubit by measuring that position and fixes its state
    \item Calls the mark() function
\end{itemize}
\subsection{STEP-5}
\textit{Quantum move: }
qmove()
\begin{itemize}
    \item Stores the inputs in control and target lists
    \item Prints `The control and target positions have been entangled'
\end{itemize}
\subsection{STEP-6}
\textit{Removing the trapped stones: }
remove()
\begin{itemize}
    \item Checks all the stones after each step and removes the particular stone that has no liberty.
\end{itemize}

\subsection{STEP-7}
\textit{Printing board after each step:}
mark()

\begin{itemize}
    \item Re-simulates the circuit
    \item Prints the board as per moves made in respective turns by the players by calling the remove() and board() functions.
    \item Returns the number of white and black stones captured
\end{itemize}

This is how cmove and qmove work in our game. This will be iterated in the subsequent moves in the game.\\

\begin{figure}
    \centering
    \subfigure[]{}
    \centering
    \includegraphics[scale=0.06]{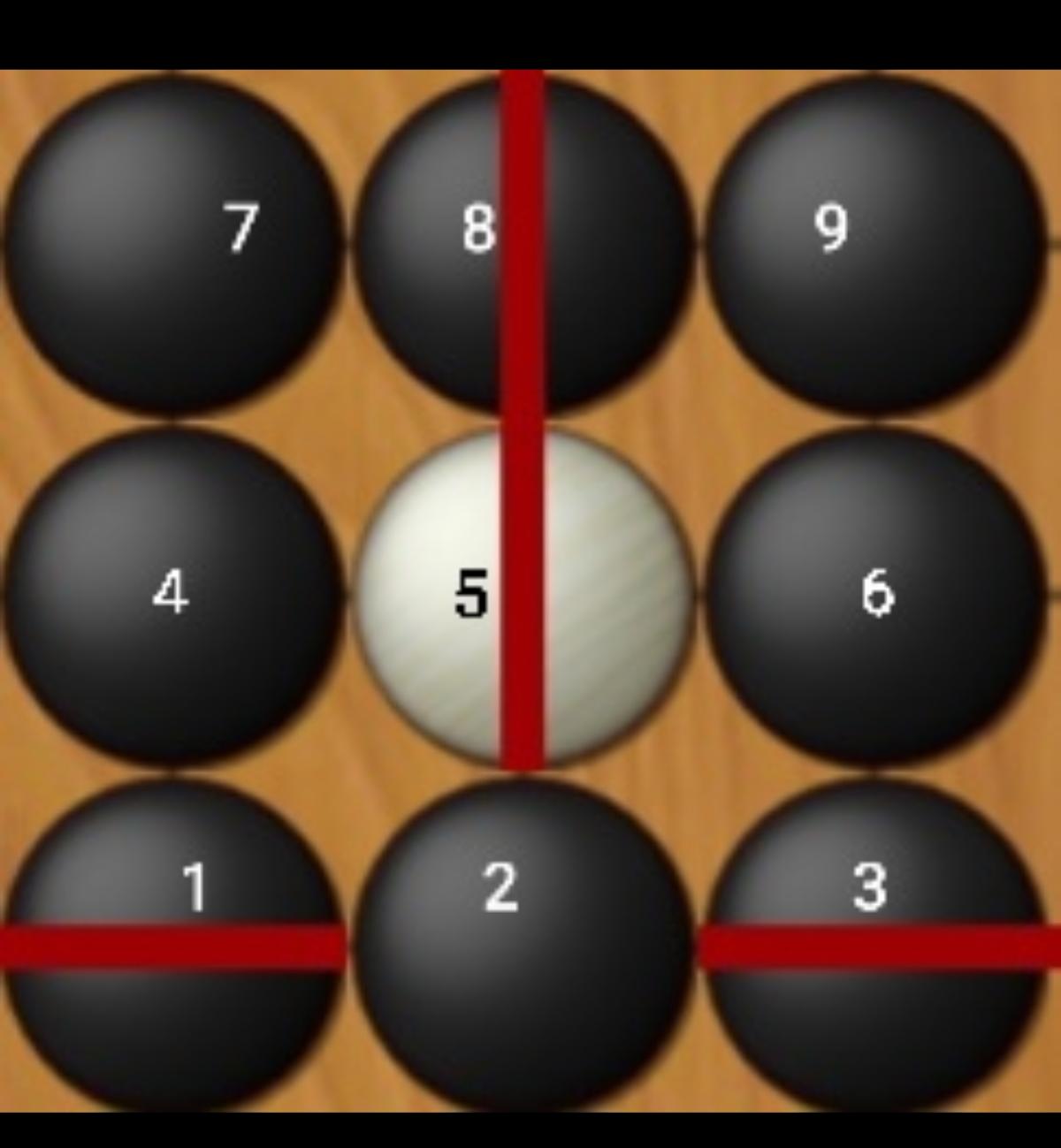}
    \subfigure[]{}
    \centering
    \includegraphics[scale=0.06]{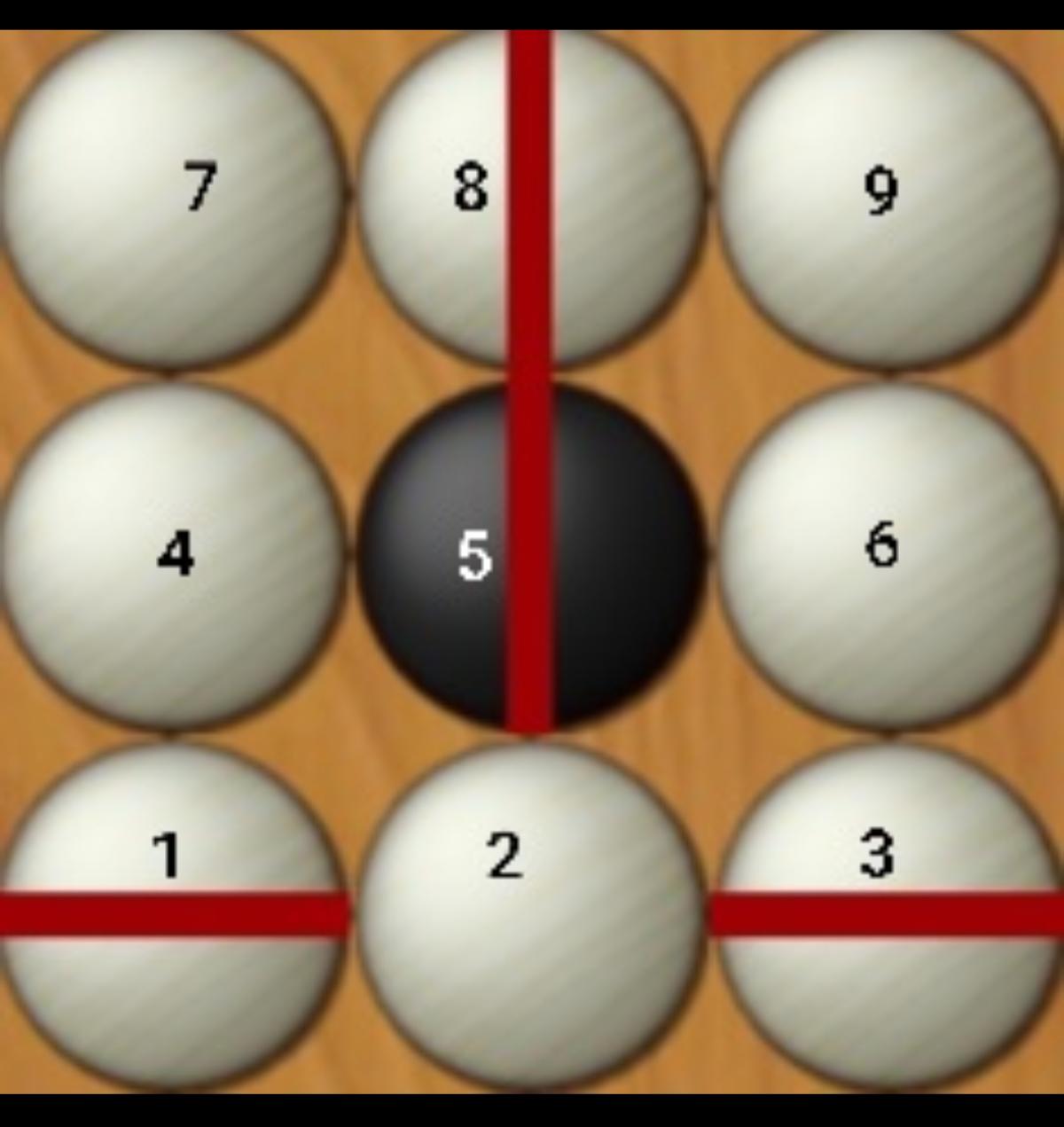}
    \caption{Here in the first figure there is a general visualization of white captured stone in $3 \times 3$ board. The second one is the visualization of black captured stone in $3\times3$ board. The positions are mentioned here.}
    \label{FF}
\end{figure}

\textbf{\MakeUppercase{Quantum Circuit for Capturing}}\\
\\
Now we proceed to do the capturing parts. First of all, let us take a $3\times3$ board for simplicity. Because it is hard to understand the capturing part in the $4\times4$ board directly. In $3\times3$ board, we can see there are 9 positions. Now, we can observe in Fig. \ref{FF}, where we have to take the following conventions.

\begin{itemize}
\item[(i)] For capturing of black :- Result is 0.
\item[(ii)] For capturing of white :- Result is 1.
\end{itemize}

Positions:- See Fig. \ref{FF}.

\begin{figure}
\centering
\includegraphics[scale=0.09]{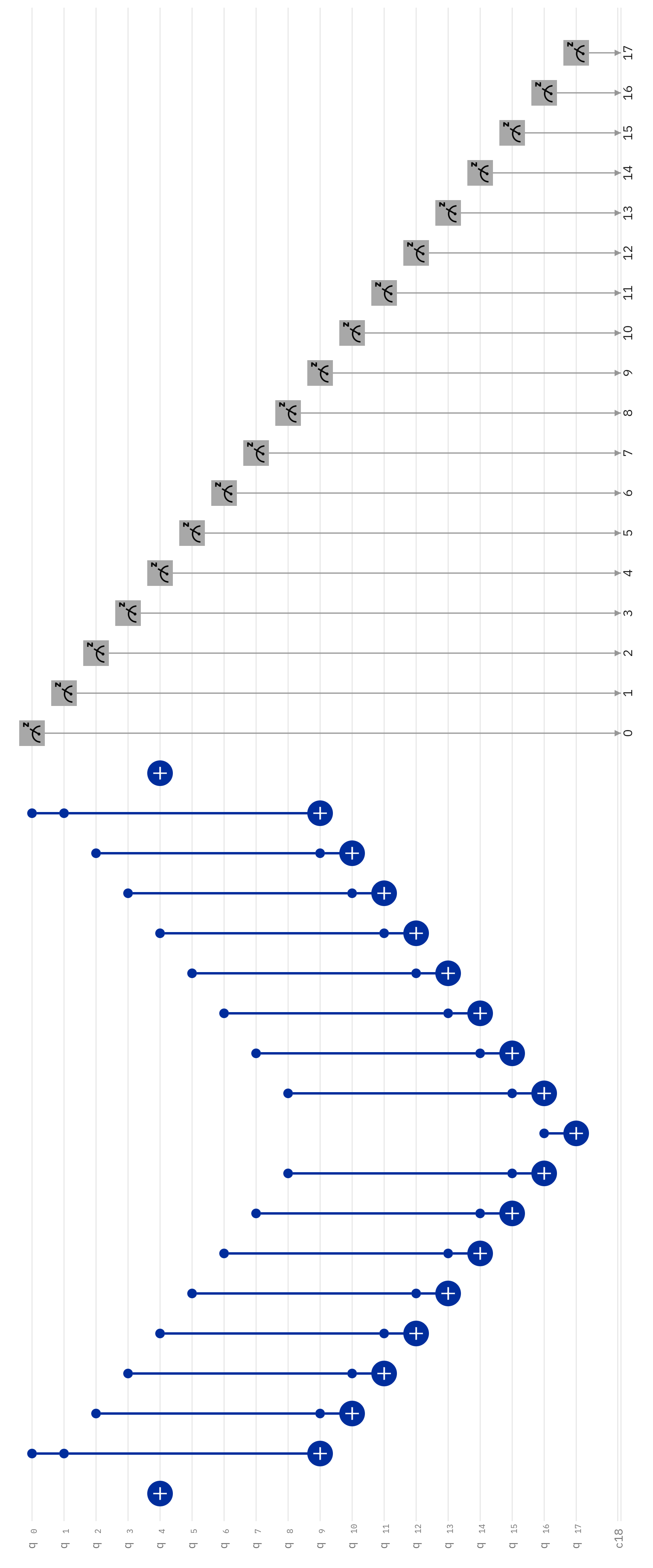}
\caption{Circuit explanation for capturing of $3 \times 3$ board at position no. 5. Here we took 18 qubits, 8 work qubits for one capturing and 9 qubits for 9 positions of the board. And 1 is the target qubit for capturing position. Like this we can make the circuit of $4 \times 4$ board at different positions. Here we need 28 qubits. There is a generalised formula for $n \times n$ board in the describe paragraph.}
\label{CC}
\end{figure}

The capturing of both the white and black stones is presented in Fig. \ref{FF}. The captured position is $5{th}$ (Fig. \ref{FF}) in both cases. We take 9 qubits for each and every position for $3\times3$ board. Now we use comparator for the stones. We are comparing the simulated results of each position with the simulates result of captured position. That means for black capture we have result 0 in the $5^{th}$ qubit and others have result 1 and vice versa for white stone. As a result, we have to use control qubits for 8 positions. Therefore, we have to take work qubits i.e., n-1 [where `n' is the number of control qubits]. According to formula n-1 we have to take 8 work qubits and one target qubit. Hence, total and others have result 1 and vice versa for white stone. As a result, we have to use control qubits for 9 positions. Therefore, we have to take work qubits i.e. n-1 (where `n' is the number of control qubits). According to formula n-1 we have to take 8 work qubits and one target qubit. Hence, total number of qubits is $9+8+1 = 18$. We give the circuit diagram for the capturing (Fig. \ref{CC}).

\begin{figure*}%
\centering
\subfigure[][]{%
\label{qgo_Fig3}%
\includegraphics[width=0.3\linewidth, height=2.75in]{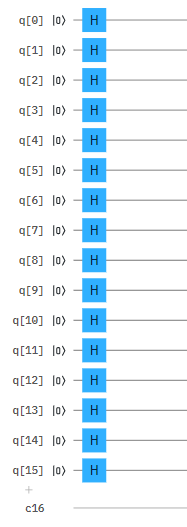}}%
\hspace{36pt}%
\subfigure[][]{%
\label{qgo_Fig6}%
\includegraphics[width=0.3\linewidth,height=2.75in]{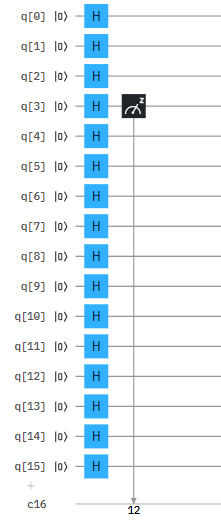}} \\
\subfigure[][]{%
\label{qgo_Fig4}%
\includegraphics[width=0.3\linewidth,height=2.75in]{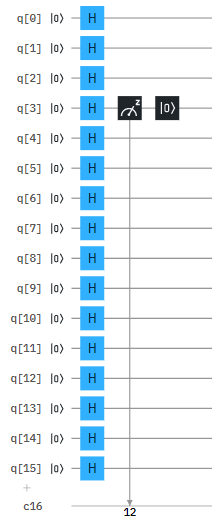}}%
\hspace{36pt}%
\subfigure[][]{%
\label{qgo_Fig7}%
\includegraphics[width=0.3\linewidth,height=2.75in]{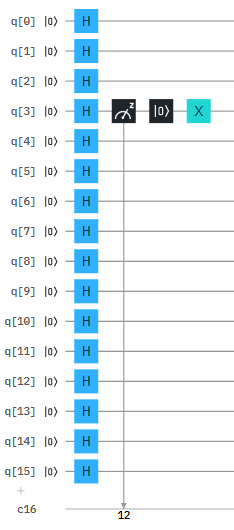}}\\
\subfigure[][]{%
\label{qgo_Fig5}%
\includegraphics[width=0.25\linewidth, height=0.75in]{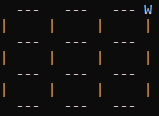}}%
\hspace{64pt}%
\subfigure[][]{%
\label{qgo_Fig8}%
\includegraphics[width=0.25\linewidth,height=0.75in]{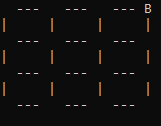}}%
\caption[Explanation]{Circuit explanation and output after first move:
\subref{qgo_Fig3} the basic circuit with 16 qubits and 16 classical bits for 4$\times$4 board;
\subref{qgo_Fig6} measuring the first chosen box (here 4 i.e. q3);
\subref{qgo_Fig4} adding Reset gate if the measured value is $\Ket{0}$; 
\subref{qgo_Fig7} adding a Reset and a Not gate if the measured value is $\Ket{1}$;
\subref{qgo_Fig5} Output if measured value is $\Ket{0}$; and,
\subref{qgo_Fig8} Output if measured value is $\Ket{0}$}%
\label{fig:ex3}%
\end{figure*}

\begin{figure}[!h]
\centering
\subfigure[]{\label{qgo_Fig9}}
\centering
\includegraphics[width=0.3\textwidth]{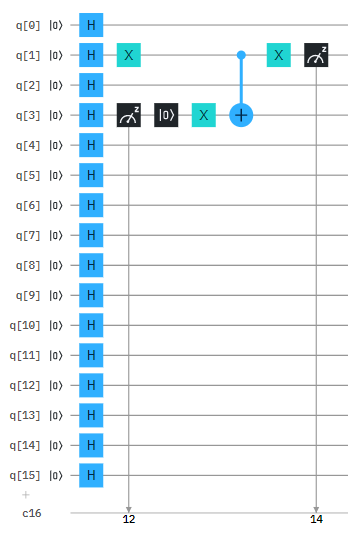} \\
\subfigure[]{\label{qgo_Fig10}}
\centering
\includegraphics[width=0.2\textwidth]{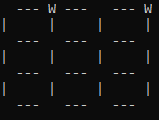}
\subfigure[]{\label{qgo_Fig11}}
\centering
\includegraphics[width=0.2\textwidth]{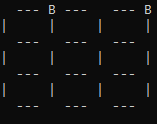}
\caption{Circuit explanation and output after first move: \subref{qgo_Fig9} this circuit if $q3$ collapses to $\Ket{1}$ and anti-CNOT gate is applied when White chooses quantum move to entangle it with the first qubit i.e., second box, \subref{qgo_Fig10} this is the output if $q1$ collapses to $\Ket{0}$; $q3$ changes its state, and \subref{qgo_Fig11} this is the output if $q1$ collapses to $\Ket{1}$; $q3$ remains in $\Ket{1}$ state.}
\end{figure}

\begin{figure}[!h]
\centering
\includegraphics[scale=0.06]{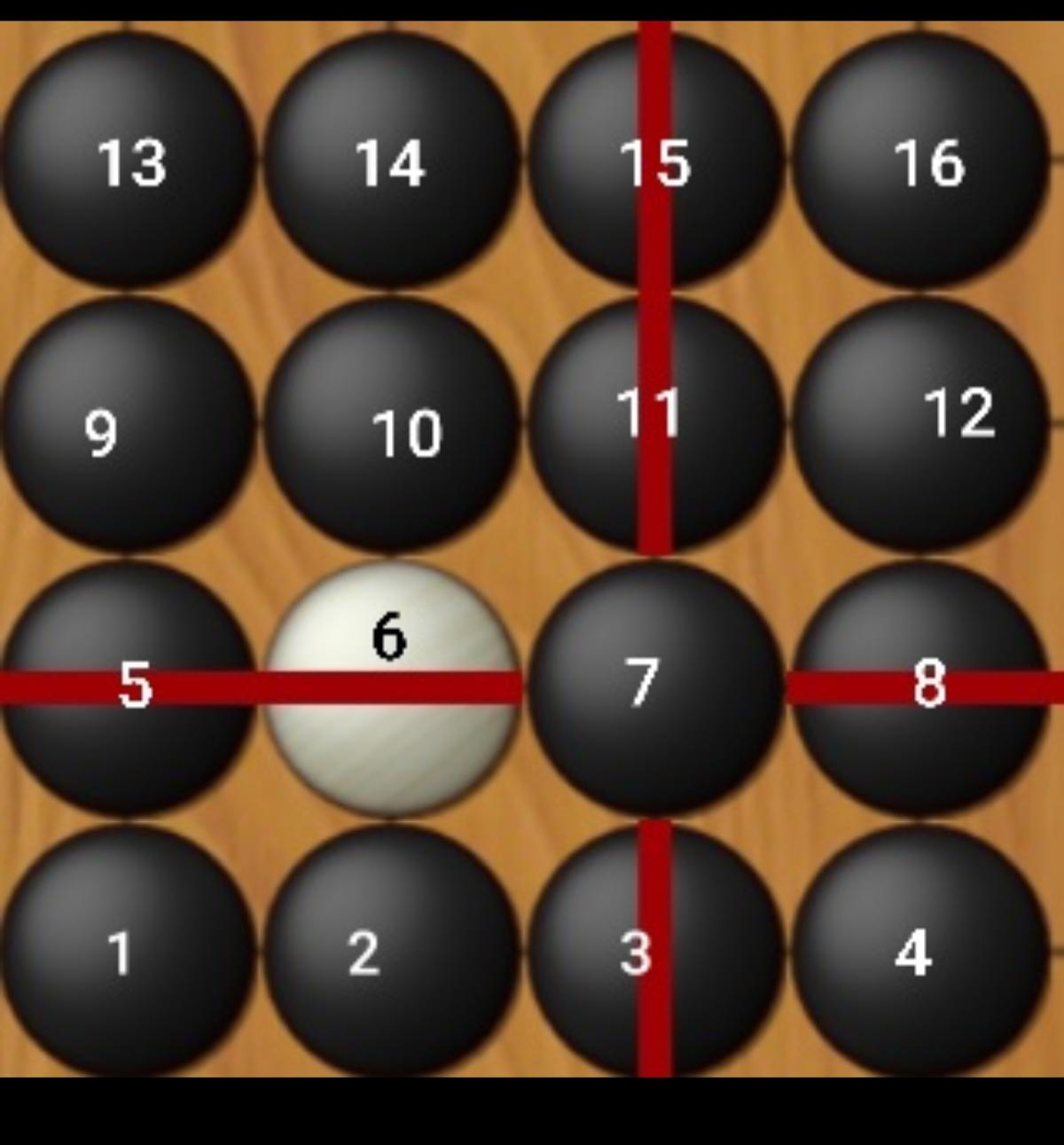}
\caption{Fig for capturing of $4\times4$ board at position number 6. Similarly, we can see capturing at positions of 7, 10 and 11.}
\label{FG}
\end{figure}
Similarly, we can proceed with the $4\times4$ board. There are 16 positions in the $4\times4$ board. That means we have to take 16 qubits for this board. we need $9$ positions for one capturing (Fig. \ref{FG}). So, here also we have to take $8$ work qubits. Clearly we can see $4$ captures are possible. So again we have to take $4$ target qubits. As a result total number of qubits is $16+8+4 = 28$. In this case, we have to take $8$ control gates and one anti-control gate. We can make circuit by looking the circuit diagram of $3\times3$ board. In general we can take $n\times n$ board and perform same operation that we have done for $3\times3$ board and $4\times4$ board. Previously, we have seen that total number of position is $n^2$. However, for capturing of one position we always need $9$ positions, so $8$ work qubits are needed for one capturing. For $n\times n$ board there are $(n-2)^2$ number of captures possible. But we use $8$ CNOT gates and $1$ anti-control gate for $1$ capturing. And we also need $(n-2)^2$ number of target qubits. Therefore, total number of qubits for $1$ capturing in $n\times n$ board is $n^2 + (n-2)^2 + 8$.

\section{Quantum Metaphors}
Quantum Go game adds a level of complexity by allowing players to explore the possibilities coming from each position being $\Ket{0}$ and $\Ket{1}$ at once. This phenomenon simulates quantum superposition, which is the principle that states physical objects need not have a definite attribute. In quantum theory, physical systems exhibiting superposition are studied through a quantum measurement. This can be interpreted as a phenomenon of quantum collapse, where the physical states of a system in superposition are reduced to states which are no longer in a superposition. In quantum Go game, an analogous process occurs when a classical move is applied on a qubit in superposition. Another quantum phenomenon addressed to is entanglement: when the quantum move is applied on the two qubits (control and target), the quantum state of each qubit cannot be described independently until collapse has occurred through the classical move. And after one qubit is measured, the state of the other does not require measurement; knowing one qubit's state defines the other. We can illustrate these ideas concerning the previous example more precisely.

\begin{itemize}
\item \textbf{Superposition:} In the beginning, all the qubits are in a superposition of $\Ket{0}$ and $\Ket{1}$ with a $50\%$ probability each.
\item \textbf{Collapse:} In the first step, 3rd qubit is measured and its state is now well-defined.
\item \textbf{Entanglement:} In the next step, first and third qubits are entangled by the quantum move. In the later step, on applying anti-control Not gate, the states of both qubits were found to be the same.
\end{itemize}

\section{Conclusion}
Due to the introduction of the probabilistic approach in designing of the game, the naive first impression might be that the game is completely based on luck. However, it is not fully right. Firstly, we all know that probability is inherent in the quantum phenomena. So, in the quantum Go game, the concept of probability is inevitable. Secondly, as the players start playing the game, they will come to realize that we need a proper strategy to follow to be the winner. You need to be strategic during quantum moves while you entangle two boxes. The luck factor is handy only in the case of classical moves. We can note that a good strategy would be not to entangle a box with another which is already in your favored state. Let us illustrate it in the following example: Bob goes for a classical move to collapse any of the boxes, and it collapses to $\Ket{0}$. So the box turns `W' now and if in the next move, Alice entangles this, and the control of this move in any subsequent step collapses to $\Ket{0}$ which makes the previously `W' marked box `B'. Strategies should be made before classical moves also. The players need to foresee what effect collapsing of a particular box to either of the states may have on his position in the game. Comparing with the classical Go, this element of probability in quantum Go adds a more exciting feature to the game. It requires more detailed thinking. Depending on the board size, the number of qubits needed is $(n^2)$ and accordingly, needed are $(n^2)$ classical bits to measure them. For each step, the time complexity is proportional to the number of gates involved. For the same board size, its classical counterpart has $O(n^2)$ time complexity.\\
Though some of the interesting quantum phenomena such as complementarity, non-locality, and contextuality have no analogs in quantum Go, it is a fun way to get an insight into a few fundamental quantum phenomena.

\end{document}